\documentclass[prb,twocolumn,showpacs,superscriptaddress]{revtex4}

\usepackage{amsmath}
\usepackage{amssymb}
\usepackage{amsthm}
\usepackage{graphicx}

\newcommand{\tr}{\operatorname{tr}}
\newcommand{\altproj}{\operatorname{Alt}}

\begin{document}
\bibliographystyle{apsrev}

\title{Statics and Dynamics of Quantum $XY$ and Heisenberg Systems on Graphs}

\author{Tobias J.\ Osborne}
\email[]{Tobias.Osborne@rhul.ac.uk} \affiliation{School of
Mathematics, University of Bristol, University Walk, Bristol BS8
1TW, United Kingdom}\affiliation{Department of Mathematics, Royal
Holloway University of London, Egham, Surrey TW20 0EX, United
Kingdom}

\date{\today}

\begin{abstract}
We consider the statics and dynamics of distinguishable spin-$1/2$
systems on an arbitrary graph $G$ with $N$ vertices. In particular,
we consider systems of quantum spins evolving according to one of
two hamiltonians: (i) the $XY$ hamiltonian $H_{XY}$, which contains
an $XY$ interaction for every pair of spins connected by an edge in
$G$; and (ii) the Heisenberg hamiltonian $H_{\text{Heis}}$, which
contains a Heisenberg interaction term for every pair of spins
connected by an edge in $G$. We find that the action of the $XY$
(respectively, Heisenberg) hamiltonian on state space is equivalent
to the action of the adjacency matrix (respectively, combinatorial
laplacian) of a sequence $G_k$, $k=0,\ldots, N$ of graphs derived
from $G$ (with $G_1=G$). This equivalence of actions demonstrates
that the dynamics of these two models is the same as the evolution
of a free particle hopping on the graphs $G_k$. Thus we show how to
replace the complicated dynamics of the original spin model with
simpler dynamics on a more complicated geometry. A simple corollary
of our approach allows us to write an explicit spectral
decomposition of the ${XY}$ model in a magnetic field on the path
consisting of $N$ vertices. We also use our approach to utilise
results from spectral graph theory to solve new spin models: the
$XY$ model and heisenberg model in a magnetic field on the complete
graph.
\end{abstract}


\maketitle

Understanding the static and dynamic properties of interacting
quantum spins is a central problem in condensed-matter physics. A
handful of extremely powerful techniques have been developed to
tackle this difficult problem. Amongst the most well-known are the
Bethe ansatz method \cite{bethe:1931a}, Jordan-Wigner fermion
transformations \cite{jordan:1928a}, and ground-state ansatz
methods, for example, methods based on finitely correlated
states/matrix product states \cite{fannes:1994a, schollwoeck:2005a}.

The techniques developed to solve interacting many-body quantum
systems have led to the discovery of many new intriguing
nonclassical phenomena. A canonical example is the discovery of
quantum phase transitions \cite{sachdev:1999a, auerbach:1994a},
phase transitions which occur in the ground state --- a pure state
--- which are driven by quantum rather than thermal fluctuations.
However, these techniques can typically only be applied to systems
which possess a great deal of symmetry. Hence it is extremely
desirable to develop new approaches that can be applied in more
general situations.

There is a superficial similarity between the mathematics of
distinguishable quantum spins and the \emph{spectral theory of
graphs} \cite{biggs:1993a, cvetkovic:1995a, chung:1997a}, which
pertains to the dynamics of a single quantum particle hopping on a
discrete graph. In both cases there is a graph structure and a
notion of locality. In the case of graphs, locality can be
characterised by the \emph{support} of the particle wavefunction,
i.e., the position of the quantum particle. A localised particle
remains, for small times, approximately localised. (There is a
natural UV cutoff given by the graph structure, hence there is a
resulting bound on the propagation speed of the particle.) In the
case of spin systems the notion of locality emerges in the
Heisenberg picture where the support of \emph{operators} takes the
role of defining local physics. Under dynamics local operators
remain approximately local for short times. (This locality result is
a consequence of the \emph{Lieb-Robinson bound} \cite{lieb:1972a,
hastings:2004a, nachtergaele:2005a, hastings:2005b}.) In both the
case of quantum spin systems and spectral graph theory we are
interested in the eigenvalues and the eigenvectors of the generator
of time translations: the hamiltonian for the spin system; and the
adjacency matrix in the case of graphs.

It is tempting to think that the connection between quantum spin
systems and the spectral theory of graphs might be exploited in a
strong way to use the extensive spectral theory \cite{biggs:1993a,
cvetkovic:1995a, chung:1997a} developed to study finite graphs to
study quantum spin systems. However, this is not trivial. The
principle problem is that the hilbert space of a graph $G$ with $N$
vertices is given by $\mathbb{C}^N$, and the hilbert space of a spin
system with a spin-$1/2$ particle attached to each vertex of the
same graph $G$ is $\mathbb{C}^{2^N}$, which is exponentially larger.

In this paper we introduce a new way to understand the statics and
dynamics (i.e.\ the eigenvectors and eigenvalues) of a large class
of interacting spin systems. In particular, we show how to
understand the statics and dynamics of a system of $N$ spin-$1/2$
particles interacting according to the pairwise $XY$ or Heisenberg
interactions on a graph $G$ in terms of the structure of $G$. We
show that the action of the hamiltonian for the spin system is
identical to the action of an \emph{adjacency matrix} for a disjoint
union of graphs $G_k$ related to $G$. Thus we provide a direct
connection to the spectral theory of graphs for these models.

The central idea underlying this paper is that complicated physics
of a system of distinguishable spin-$1/2$ particles interacting
pairwise on a simple geometry given by a graph $G$ are
\emph{equivalent} to the sometimes simpler physics of a
\emph{single} free spinless particle hopping on a much complicated
graph $\mathcal{G}$ (which is a disjoint union of the graphs $G_k$).
This equivalence can be exploited in certain situations to extract
partial and sometimes complete information about the statics and
dynamics of $H$.

The outline of this paper is as follows. We begin by reviewing how
the action of $H_{XY}$ and $H_{\text{Heis}}$ breaks the hilbert
space into a direct sum of subspaces $\Gamma^k$, $k=0, \ldots, N$.
We then consider the action of the hamiltonians on each closed
subspace separately. We define a new type of graph product, the
\emph{graph wedge product} $G\wedge G$. We show that the action of
$H_{XY}$ (respectively, $H_{\text{Heis}}$), when restricted to
$\Gamma^k$ is the same as that of the adjacency matrix
(respectively, combinatorial laplacian) of $\bigwedge^k{G}$. We then
show that we can diagonalise the adjacency matrix for the graph
$\bigwedge^k{G}$ to obtain the eigenvalues and eigenvectors for the
spin system. In the case of the path on $N$ vertices this allows us
to write out an explicit specification of the eigenvalues and
eigenvectors of the $XY$ model in a magnetic field on the line.
Finally, we use our connection to use results from spectral graph
theory to solve the $XY$ and Heisenberg models on the complete graph
$K_N$ on $N$ vertices.

Let us begin by defining the main objects of our study. We start
with a little graph-theoretic terminology: let $G = (V,E)$ be a
graph, that is, a finite set $V$ of \emph{vertices} and a collection
$E$ of $2$-element subsets of $V$ called \emph{edges}. We fix a
labelling $\ell(v)\in\{0,\ldots,N-1\}$ of the vertices once and for
all. This labelling induces an ordering of the vertices which we
write as $v_j>v_k$ if $\ell(v_j) > \ell(v_k)$. In the following we
will not refer to the labelling explicitly, only implicitly via this
ordering. The \emph{degree} $d_v$ of a vertex $v$ is equal to the
number of edges which have $v$ as an endpoint. The \emph{adjacency
matrix} $[A(G)]_{v,w}$ for $G$ is the $\{0,1\}$-matrix of size
$|V(G)|\times |V(G)|$ which has a $1$ in the $(v,w)$ entry if there
is an edge connecting $v$ and $w$. Finally, we define the hilbert
$\mathcal{H}_G$ space of the graph $G$ to be the vector space over
$\mathbb{C}$ generated by the orthonormal vectors $|v\rangle$,
$\forall v\in V(G)$, with the canonical inner product $\langle
v|w\rangle =\delta_{v,w}$.

We consider $N=|V|$ \emph{distinguishable} spin-$1/2$ subsystems
interacting according to the following two hamiltonians: the $XY$
interaction
\begin{equation}\label{eq:xyham}
H_{XY} = \frac{1}{2}\sum_{v\sim w}
({\sigma}^x_v{\sigma}^x_w+{\sigma}^y_v{\sigma}^y_w),
\end{equation}
and the Heisenberg-interaction
\begin{equation}\label{eq:heisham}
H_{\text{Heis}} = -\frac{1}{2}\sum_{v\sim w}
(\boldsymbol{\sigma}_v\cdot\boldsymbol{\sigma}_w-I_vI_w),
\end{equation}
where $\boldsymbol{\sigma}\triangleq\left((\begin{smallmatrix}0&1\\
1&0\end{smallmatrix}),(\begin{smallmatrix}0&-i\\
i&0\end{smallmatrix}),(\begin{smallmatrix}1&0\\
0&-1\end{smallmatrix}) \right)$ is the usual vector of Pauli
operators, $I_v$ is the identity operator acting on the
tensor-product subspace $\mathcal{H}_G$ associated with vertex $v$,
and $v$ and $w$ are vertices of the graph where $\sum_{v\sim w}$
means that we sum exactly once over all vertices such that there is
an edge connecting $v$ and $w$. (We can, with very little extra
effort, consider an additional constant magnetic field in the $z$
direction. For simplicity we ignore this at the moment. We outline
how to include such fields toward the end of this paper.)

It is a well-known property of the $XY$ and Heisenberg interaction
\cite{auerbach:1994a} that they commute with the total $z$-spin
operator $S^z\triangleq\sum_{v\in V}\sigma^z_v$,
\begin{equation}
[S^z, {\sigma}^x_v{\sigma}^x_w+{\sigma}^y_v{\sigma}^y_w]=0,
\quad\forall v,w\in V,
\end{equation}
and
\begin{equation}
[S^z, \boldsymbol{\sigma}_v\cdot\boldsymbol{\sigma}_w]=0,
\quad\forall v,w\in V.
\end{equation}
(Indeed, the Heisenberg interaction commutes with the total $x$- and
$y$-spin operators as well, so it is invariant under the action of
$\text{\emph{SU}}(2)$.) In this way we see that the action of
$H_{XY}$ and $H_{\text{Heis}}$ breaks the hilbert space
$\mathcal{H}\cong\mathbb{C}^{2^N}$ into a direct sum
$\mathcal{H}\cong \bigoplus_{k=0}^N \Gamma^k$, where
\cite{endnote15}
\begin{equation}
\begin{split}
\Gamma^0 &= \{|00\cdots0\rangle\} \\
\Gamma^1 &= \{|10\cdots0\rangle, |01\cdots0\rangle, \ldots,
|00\cdots1\rangle\} \\
\Gamma^2 &= \{|11\cdots0\rangle, |101\cdots0\rangle, \ldots,
|0\cdots11\rangle\} \\
&\vdots \\
\Gamma^N &= \{|11\cdots1\rangle\}, \\
\end{split}
\end{equation}
are the vector spaces of total spin $\tr(S^z P_{\Gamma^k}) = k$ and
dimension $\dim\Gamma^k = \binom{N}{k}$. ($P_{\Gamma^k}$ denotes the
projector onto $\Gamma^k$.)

We now describe a fundamental connection between the vector spaces
$\Gamma^k$ and exterior vector spaces. To do this we define the
following vector spaces $\bigwedge^k(\mathcal{H}_G)$. Firstly,
$\bigwedge^0(\mathcal{H}_G)\triangleq \mathbb{C}$ and
$\bigwedge^1(\mathcal{H}_G)\triangleq \mathcal{H}_G$. For general
$k$ we define $\bigwedge^k(\mathcal{H}_G)$ to be the vector space
$\bigotimes_{l=0}^{k-1}\mathcal{H}_G$ modulo the vector space
$\mathcal{A}_G$ generated by all elements of the form $|v_0,
v_1,\ldots, v_{k-1}\rangle$ where $|v_j\rangle\in \mathcal{H}_G$ and
where $|v_j\rangle=|v_{j'}\rangle$ for some $j\not= j'$. Another way
of saying this is that the exterior vector space
$\bigwedge^k(\mathcal{H}_G)$ is spanned by vectors $|v_0,
v_1,\ldots, v_{k-1}\rangle$ where no two $|v_j\rangle$,
$|v_k\rangle$, $\forall j\not=k$, are the same.

To write a basis for $\bigwedge^k(\mathcal{H}_G)$ we need to
introduce the wedge product $\wedge:\mathcal{H}_G\times
\mathcal{H}_G\times\cdots\times\mathcal{H}_G\rightarrow
\bigotimes_{l=0}^{k-1}\mathcal{H}_G$ which is defined by
\begin{equation}
|v_{0}\wedge v_{1}\wedge\cdots \wedge v_{k-1}\rangle \triangleq
\frac{1}{k!}\sum_{\pi\in S_k} \epsilon(\pi) |v_{\pi(0)},
v_{\pi(1)}, \ldots, v_{\pi(k-1)}\rangle,
\end{equation}
where $S_k$ is the symmetric group on $k$ letters and
$\epsilon(\pi)$ is the \emph{sign} of the permutation $\pi$. Note
that $\dim\bigwedge^k(\mathcal{H}_G) = \binom{N}{k}$.

A basis for $\bigwedge^k(\mathcal{H}_G)$ is given by the vectors
$|v_{0} \wedge v_{1} \wedge \cdots \wedge v_{k-1}\rangle$ with
$v_{j}\in V(G)$ and $v_{k-1}>v_{k-2}>\cdots
>v_{0}$. (For a quick review of exterior vector spaces
and exterior algebras see \cite{fulton:1991a} or, for a more
leisurely treatment, see \cite{lang:2002a}.)

We can now make manifest the promised connection between
$\Gamma^k$ and $\bigwedge^k(\mathcal{H}_G)$. Because these two
families of vector spaces have the same dimension we see they are
immediately isomorphic as vector spaces over $\mathbb{C}$. We
identify the state $|1_{v_0}, 1_{v_1}, \ldots, 1_{v_{k-1}}\rangle
\in \Gamma^k$ which has a $1$ at positions/vertices $v_{k-1}>
v_{k-2} > \cdots
> v_0$ and zeros elsewhere, with the basis vector $|v_{0}\wedge v_1\wedge\cdots\wedge
v_{k-1}\rangle\in \bigwedge^k(\mathcal{H}_G)$.

We now turn to the definition of our graph product, the
\emph{graph wedge product}. We define the graph wedge product
$\bigwedge^k G$ of a graph $G$ to be the graph with vertex set
$V(\bigwedge^k G) \triangleq \{ (v_{0}, v_{1}, \ldots, v_{k-1})
\,|\, v_{j}\in V(G), v_{k-1}> v_{k-2} > \cdots
> v_0 \}$. We write
vertices of $\bigwedge^k G$ as $v_{0}\wedge v_{1}\wedge\cdots\wedge
v_{k-1}$. We connect two vertices $v_{0}\wedge
v_{1}\wedge\cdots\wedge v_{k-1}$ and $w_{0}\wedge
w_{1}\wedge\cdots\wedge w_{k-1}$ in $\bigwedge^k G$ with an edge if
there is a permutation $\pi\in S_k$ such that $w_{j} = v_{\pi(j)}$
for all $j=0, \ldots, k-1$ except at one place $j=\pi(l)$ where
$(v_{l}, w_{\pi(l)})\in E(G)$ is an edge in $G$. Obviously the
hilbert space $\mathcal{H}_{\bigwedge^k G}$ of the graph
$\bigwedge^k G$ is isomorphic to $\bigwedge^k(\mathcal{H}_G)$.

The $n$-fold graph wedge product of a graph $G$, namely $G_n =
\bigwedge^n G$, has been studied in the graph theory literature
where it has been referred to as an \emph{$n$-tuple vertex graph}
\cite{alavi:1991a, alavi:2002a}. There appears to be no literature
on the spectral properties of $n$-tuple vertex graphs in general:
the results of this paper appear to provide the first investigation
of the spectral properties of such graphs.

The adjacency matrix for $\bigwedge^kG$ can be found via the
following procedure. Let $M\in\mathcal{B}(\mathcal{H}_G)$ be a
linear operator from $\mathcal{H}_G$ to $\mathcal{H}_G$. (We are
using the symbol $\mathcal{B}(\mathcal{H}_G)$ to denote the vector
space of all bounded operators on $\mathcal{H}_G$.) Define the
operation $\triangle^k:\mathcal{B}(\mathcal{H}_G)\rightarrow
\bigotimes_{j=0}^{k-1}\mathcal{B}(\mathcal{H}_G)$ by
\begin{equation}\label{eq:coprod2}
\triangle^k(M) \triangleq \sum_{j=0}^{k-1} I_{01\cdots j-1}\otimes
M_j \otimes I_{j+1\cdots k-1}.
\end{equation}

Define also the projection $\altproj :
\bigotimes_{j=0}^{k-1}\mathcal{H}_G \rightarrow
\bigwedge^k{\mathcal{H}_G}$ by
\begin{multline}\label{eq:projadj}
\altproj|\psi_0, \psi_1, \ldots, \psi_{k-1}\rangle \triangleq \\
\frac{1}{k!}\sum_{\pi\in S_k} \epsilon(\pi) \pi\cdot (|\psi_0,
\psi_1, \ldots, \psi_{k-1}\rangle),
\end{multline}
where the action of the symmetric group $S_k$ is defined via
$\pi\cdot(|\psi_0, \psi_1, \ldots, \psi_{k-1}\rangle)
\triangleq|\psi_{\pi(0)}, \psi_{\pi(1)}, \ldots,
\psi_{\pi(k-1)}\rangle$. One can verify, with a little algebra,
that the specification Eq.~(\ref{eq:projadj}) of $\altproj$ is
well defined.

Using $\triangle^k$ and the projection $\altproj$ we construct the
following matrix which encodes the structure of $\bigwedge^k G$:
\begin{equation}
C\left(\bigwedge^k G\right) = \altproj \triangle^k(A(G))\altproj.
\end{equation}
It is easily verified that $\langle v_0\wedge v_1\wedge\cdots\wedge
v_{k-1}|C\left(\bigwedge^k G\right) |w_0\wedge w_1\wedge \cdots
\wedge w_{k-1}\rangle = \pm1$ if and only if $v_j=w_j$ for all $j$
except at exactly one place $j=l$, where $(v_l,w_l)\in E(G)$. All
the other entries are zero. (We are exploiting Dirac notation to
write the matrix elements $[M]_{v,w}$ of a matrix $M$ as $\langle
v|M|w\rangle$.) The adjacency matrix $A\left(\bigwedge^k G\right)$
of $\bigwedge^k G$ is typically different from $C\left(\bigwedge^k
G\right)$ and is found by replacing all instances of $-1$ with $1$
and leaving the zero entries alone.

Suppose we know the complete spectral decomposition for $A(G)$,
i.e.\ we know a specification of both the eigenvalues $\lambda_j$
and eigenvectors $|\lambda_j\rangle$ in
\begin{equation}
A(G) = \sum_{j=0}^{N-1} \lambda_j |\lambda_j\rangle\langle
\lambda_j|.
\end{equation}
We apply Eq.~(\ref{eq:coprod2}) to write
\begin{multline}
\triangle^k (A(G))=\\ \sum_{j_0, j_1, \ldots, j_{k-1} = 0}^{N-1}
\mu_{j_0,j_1,\ldots,j_{k-1}} |\lambda_{j_0}, \ldots,
\lambda_{j_{k-1}}\rangle\langle \lambda_{j_0}, \ldots,
\lambda_{j_{k-1}}|,
\end{multline}
where
\begin{equation}\label{eq:anteval}
\mu_{j_0,j_1,\ldots,j_{k-1}}\triangleq\lambda_{j_0}+\lambda_{j_1}+\cdots+\lambda_{j_{k-1}}.
\end{equation}

Now consider the action of the projector $\altproj$ on the vector
$|\lambda_{j_0},  \ldots, \lambda_{j_{k-1}}\rangle$:
\begin{equation}
\altproj|\lambda_{j_0},  \ldots, \lambda_{j_{k-1}}\rangle =
\frac{1}{k!}\sum_{\pi\in S_k} \epsilon(\pi)\pi\cdot(|\lambda_{j_0},
 \ldots, \lambda_{j_{k-1}}\rangle).
\end{equation}
We write the coefficients of the eigenvector $|\lambda_j\rangle$
in the basis formed by the vertices of $G$: $|\lambda_j\rangle =
\sum_{l=0}^{N-1} {\omega_{j}}^{l} |v_l\rangle$. Using this
expansion we find
\begin{multline}
|\mu_{j_0,j_1,\ldots,j_{k-1}}\rangle \triangleq
\altproj|\lambda_{j_0},
\lambda_{j_1}, \ldots, \lambda_{j_{k-1}}\rangle =\\
\frac{1}{k!}\sum_{\pi\in
S_k}\epsilon(\pi)\sum_{l_0,l_1,\ldots,l_{k-1}=0}^{N-1}
\Omega_{j_{\pi(0)}, j_{\pi(1)}, \ldots,
j_{\pi(k-1)})}^{l_0,l_1,\ldots,l_{k-1}}|v_{l_0},\ldots,v_{l_{k-1}}\rangle,
\end{multline}
where $\Omega_{j_0, j_1, \ldots, j_{k-1}}^{l_0,l_1,\ldots,l_{k-1}}
=
{\omega_{j_0}}^{l_0}{\omega_{j_1}}^{l_1}\cdots{\omega_{j_{k-1}}}^{l_{k-1}}$.
Note that $|\mu_{j_0,j_1,\ldots,j_{k-1}}\rangle$ is nonzero if and
only if $j_l$ are all distinct.

If we write $|\mu_{j_0,j_1,\ldots,j_{k-1}}\rangle$ in the basis
formed from $|v_{l_0}\wedge v_{l_1}\wedge\cdots\wedge
v_{l_{k-1}}\rangle$ we find
\begin{multline}\label{eq:antevec}
|\mu_{j_0,j_1,\ldots,j_{k-1}}\rangle =
\\ \mathcal{N}(k)\sum_{l_0,l_1,\ldots,l_{k-1}=0}^{N-1}
\Omega_{j_0,j_1,\ldots,j_{k-1}}^{l_0,l_1,\ldots,l_{k-1}}|v_{l_0}\wedge
v_{l_1}\wedge\cdots\wedge v_{l_{k-1}}\rangle,
\end{multline}
Where $\mathcal{N}(k)$ is a normalisation factor. It is readily
verified that $\{ \mu_{j_0,j_1,\ldots,j_{k-1}},
|\mu_{j_0,j_1,\ldots,j_{k-1}}\rangle \,|\, j_{k-1}>\cdots>j_1>j_0\}$
is a spectral decomposition for $C\left(\bigwedge^k G\right)$ (but
not, typically, for $A\left(\bigwedge^k G\right)$).

We now show that the actions of $H_{XY}$ and $H_{\text{Heis}}$ on
the vector spaces $\bigwedge^k(\mathcal{H}_G)$ are the same as that
of the adjacency matrix and combinatorial laplacian for the graph
$\bigwedge^k G$, respectively. This is achieved in the case of
$H_{XY}$ by noting first that
$\frac{1}{2}({\sigma}^x_v{\sigma}^x_w+{\sigma}^y_v{\sigma}^y_w) =
|01\rangle_{v,w}\langle10| + |10\rangle_{v,w}\langle01|$. The action
of this operator on $|\psi\rangle = |1_{v}, 1_{v_1}, \ldots,
1_{v_{k-1}}\rangle \in \Gamma^k$ moves the $1$ at position $v$ to
$w$ if and only if there is no $1$ in the $w$ place. In this way we
see that the hamiltonian $H_{XY}$ maps the state $|\psi\rangle$ to
an equal superposition $|\eta\rangle$ of all states which are
identical to $|\psi\rangle$ except that a $1$ at a given vertex has
been moved along an edge of $e\in E(G)$ as long as there is no $1$
at the endpoint of $e$. From this observation it is easily verified
that the action of $H_{XY}$ is the same as that of $A(\bigwedge^k
G)$ on $\Gamma^k$. (A special case of this equivalence of actions
was recently noted \cite{christandl:2003a} for the $XY$ hamiltonian
acting on the subspace $\Gamma^1$.)

For the Heisenberg interaction we note that
\begin{multline}
-\frac{1}{2}(
\boldsymbol{\sigma}_v\cdot\boldsymbol{\sigma}_w-I_vI_w) =
|01\rangle_{v,w}\langle01| + |10\rangle_{v,w}\langle10| \\ -
|01\rangle_{v,w}\langle10| + |10\rangle_{v,w}\langle01|.
\end{multline}
The action of the Heisenberg hamiltonian $H_{\text{Heis}}$ is
similar to that of $-H_{XY}$. The principle difference is that the
action of $H_{\text{Heis}}$ on $|\psi\rangle = |1_{v}, 1_{v_1},
\ldots, 1_{v_{k-1}}\rangle$ maps $|\psi\rangle$ to an equal
superposition of $(-1)|\eta\rangle$ \emph{plus} $d(|\psi\rangle)$
times $|\psi\rangle$, where $d(|\psi\rangle)$ is equal to the
number of states which can be found by swapping a $1$ at any
vertex $v$ along an edge $e\in E(G)$ as long as there is no $1$ at
the endpoint of $e$. Thus, we see that the action of
$H_{\text{Heis}}$ on $\bigwedge^k G$ is the same as that of
$D(\bigwedge^k G)-A(\bigwedge^k G)$ where $D(\bigwedge^k G)$ is
the diagonal matrix with entries $[D(\bigwedge^k G)]_{v,v} =
d(v)$, the degree of $v$.

The matrix $\mathcal{L}(G)=D(G)-A(G)$ for a graph $G$ is known as
the \emph{combinatorial laplacian} for $G$ (for a review of the
combinatorial laplacian and its properties see \cite{biggs:1993a,
cvetkovic:1995a, chung:1997a}). One of the key properties of the
laplacian is that, for graphs $G$ which are discretisations of a
smooth manifold $M$, like the unit circle $S^1$, the laplacian is
the discretisation of the smooth laplacian $\nabla^2$ on $M$. In
this way, we note that the dynamics of a quantum system defined on
a graph $G$ by setting the hamiltonian $H=\mathcal{L}(G)$ is
qualitatively equivalent to the dynamics of a free particle on
$G$. (This qualitative equivalence can be made into a quantitative
statement regarding the convergence of the heat kernel of $H$ to
the continuous version \cite{chung:1997a}.)

We now illustrate our results for the graph $G=P_N$, the path on $N$
vertices. (This is the natural discretisation of the unit interval
$[0, 1]$.) In this case the adjacency matrix $A(G_1)$ for $G_1$ is
given by
\begin{equation}
A(G_1) = \begin{pmatrix} 0 & 1 & 0 & 0 & 0 & \cdots \\ 1 & 0 & 1 & 0
& 0 & \cdots
\\ 0 & 1 & 0 & 1 & 0 & \cdots \\

\vdots & \vdots & \vdots & \ddots \\  & & & 1 & 0 & 1 \\  & & & & 1
& 0\end{pmatrix}.
\end{equation}
The adjacency matrices for $\bigwedge^k G$ are given by
$C(\bigwedge^k(G))$ as they only have entries $0$ and $(+1)$. This
is because there is no way that $H_{XY}$ or $H_{\text{Heis}}$ can
induce a transition between an ordered state $|v_{0}\wedge v_{1}
\wedge\cdots\wedge v_{k-1}\rangle$ to a state which is not ordered
correctly. (I.e.\ on the path graph there is no way the hamiltonian
can swap a $1$ around another via a different path.) The graphs
arising from our construction are illustrated in Fig.~1 in the case
of the $XY$ model on $P_6$.

The eigenvalues and eigenstates for the path graph $G$ are well
known \cite{cvetkovic:1995a, biggs:1993a},
\begin{equation}
\lambda_j = -2 \cos\left(\frac{\pi (j+1)}{N+1}\right),
\end{equation}
and
\begin{equation}
|\lambda_j\rangle =
\sqrt{\frac{2}{N+1}}\sum_{l=0}^{N-1}\sin\left(\frac{\pi (j+1)
(l+1)}{N+1}\right)|v_l\rangle,
\end{equation}
where $j=0, \ldots, N-1$. Using Eq.~(\ref{eq:anteval}) and
Eq.~(\ref{eq:antevec}) we can immediately write the eigenvalues
and eigenvectors for $H_{XY}$:
\begin{equation}
\mu_{j_0,j_1,\ldots,j_{k-1}}=\sum_{l=0}^{k-1} -2
\cos\left(\frac{\pi (j_l+1)}{N+1}\right),
\end{equation}
with $j_{k-1}>j_{k-2}>\cdots>j_{0}$, and
\begin{multline}
|\mu_{j_0,j_1,\ldots,j_{k-1}}\rangle =
\mathcal{N}(k) \times \\
\sum_{l_0,l_1,\ldots,l_{k-1}=0}^{N-1} \prod_{m=0}^{N-1}
\sin\left(\frac{\pi (j_m+1) (l_m+1)}{N+1}\right)|1_{l_0}, \ldots,
1_{l_{k-1}}\rangle,
\end{multline}
where $\mathcal{N}(k)$ is a normalisation factor.

\begin{figure}
\begin{center}
\includegraphics{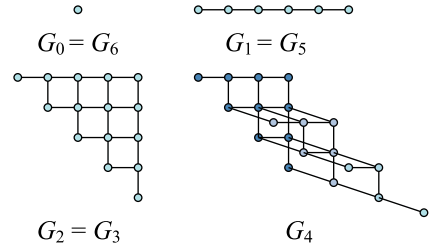}
\caption{The graphs $G_k$ that arise from our construction for the
$XY$ interaction on a path of 6 vertices. The hamiltonian $H$ for
the $XY$ model is identically equal to the adjacency matrix for the
(disconnected) graph $G_0\cup G_1 \cup \cdots \cup G_6$, i.e., $H =
A(G_0)\oplus A(G_1) \oplus \cdots \oplus A(G_6)$, where $A(G_k)$ is
the adjacency matrix of graph $G_k$. Note that $G_1=G_5$ is
equivalent to the path graph.}\label{fig:graph6}
\end{center}
\end{figure}

We now illustrate a final application of our identification of
$H_{XY}$ and $H_{\text{Heis}}$ with adjacency matrices and
laplacians for the graphs $G_k = \bigwedge^k(G)$. We consider the
$XY$ and Heisenberg models on the graph $K_N$, which is the complete
graph on $N$ vertices, meaning that every pair of vertices is
connected by an edge. In order to solve the $XY$ model on this graph
we need to understand the adjacency matrix
$A\left(\bigwedge^k(K_N)\right)$. In the case of $K_N$ the graph
$\bigwedge^k(K_N)$ is easy to describe: the vertices of
$\bigwedge^k(K_N)$ are given by $k$-subsets of $V(K_N) = \{0, 1,
\ldots, N-1\}$ and two vertices $v_j\subset V(K_N)$ and $v_k\subset
V(K_N)$ are connected if and only if they differ, as sets, in only
two places, i.e.\ $|v_j\cap v_k| = k-1$. This graph is
identical\cite{godsil:1995a} to the \emph{Johnson graph}, denoted
$J(N,k)$. The eigenvalues $\lambda_j(A(J(N,k)))$, $j = 0, 1, \ldots,
k$, of the adjacency matrix of the Johnson graph are well known and
are given by
\begin{equation}
\lambda_j(A(J(N,k))) = k(N-k) - j(N+1-j), \quad j=0,1,\ldots,k.
\end{equation}
Thus we obtain the complete spectrum for $\sigma(H_{XY}) =
\{\lambda_j(H_{XY})\,|\, j=0,1,\ldots, 2^N-1\}$:
\begin{equation}
\sigma(H_{XY}) = \bigcup_{l=0}^{N} \{\lambda_j(J(N,l))\,|\, j =
0,\ldots, l\}
\end{equation}
We note that the ground-state energy $E_0$ for $H_{XY}$ is
$E_0=-N/2$. Similarly, we observe that the degree of every vertex of
the Johnson graph $J(N,k)$ is the same, $d = k(N-k)$, so that the
eigenvalues of the Heisenberg model on the complete graph are given
by
\begin{equation}
\sigma(H_{\text{Heis}}) = \{j(N+1-j)\,|\, j=0, 1, \ldots, N\}.
\end{equation}

\begin{figure}
\begin{center}
\includegraphics{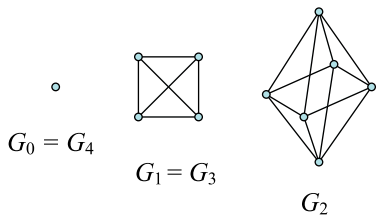}
\caption{The graphs $G_k$ that arise from our construction for the
$XY$ interaction on the complete graph $K_4$ on four vertices. The
hamiltonian $H_{XY}$ for the $XY$ model is identically equal to the
adjacency matrix for the (disconnected) graph $G_0\cup G_1 \cup
\cdots \cup G_4$. Note that $G_1 = G_3 = K_4$.}\label{fig:graphsym4}
\end{center}
\end{figure}

While it has been known how to solve the $XY$ model on the path
$P_N$ using the Jordan-Wigner transformation, the $XY$ and
Heisenberg models on the complete graph $K_N$ have defied solution
using Jordan-Wigner, Bethe ansatz, or any other method.

We note that it is straightforward to include a magnetic field
term $B S^z$ to our hamiltonians Eq.~(\ref{eq:xyham}) and
Eq.~(\ref{eq:heisham}) because it leaves the eigenvectors
unchanged and only shifts the eigenvalues according to which
subspace $\Gamma^k$ they are associated with.

Our approach should be compared with the method of Jordan and
Wigner which can be used to solve many varieties of $XY$-type
model \cite{jordan:1928a}. The Jordan-Wigner transformation
utilises exactly the same feature of the $XY$ interaction as our
method, i.e.\ that it conserves total $z$-spin. In addition, the
Jordan-Wigner transformation also (implicitly) draws a
correspondence between the vector spaces $\Gamma^k$ and
$\bigwedge^k(\mathcal{H}_G)$. However, the two methods differ when
it comes to actually calculating the eigenstates of $H_{XY}$. We
solve for the eigenvalues and eigenstates of $H_{XY}$ by
understanding the spectral properties of the associated graphs
$\bigwedge^k G$. The Jordan-Wigner method proceeds by constructing
a \emph{fermionic} hamiltonian $\widetilde{H}_{XY}$ (i.e.\ a
hamiltonian written in terms of \emph{fermionic operators} $b$ and
$b^\dag$) whose action on $\bigwedge^k(\mathcal{H}_G)$ is
isomorphic to $H_{XY}$ on $\Gamma^k$. This fermionic hamiltonian
(which is quadratic) is easily solved via Boguliubov
transformation. At this point the eigenvalues and eigenvectors can
be constructed trivially. Unfortunately, while the eigenvalues and
eigenvectors are simply specified in terms of the fermion
operators $b$ and $b^\dag$, the task of inverting the
Jordan-Wigner transformation to obtain a representation of the
eigenvectors in terms of states in the original basis is a lengthy
process.

In contrast, we have explicitly constructed the eigenstates of the
$XY$ model on a path. (And, indeed, we can construct the
eigenvalues and eigenstates for any graph $G$ for which we can
calculate the spectral decomposition of the graphs $\bigwedge^k
G$.)

In essence, the approach of Jordan and Wigner tries to understand
the dynamics of a collection of $k$ noninteracting fermions
hopping on a graph. On the other hand, our approach explicitly
constructs the \emph{configuration space} of the $k$ fermions,
reducing their dynamics to the dynamics of a single \emph{free
particle} on a (larger) graph.

Perhaps a more intriguing difference between our approach and the
Jordan-Wigner method is that our method can also be applied to the
Heisenberg interaction. The Jordan-Wigner transformation, when
applied to Heisenberg interactions, results in a highly nontrivial
nonlocal fermion hamiltonian whose solution is unknown; in the
Heisenberg case one must resort to the Bethe ansatz method
\cite{bethe:1931a}. In contrast to this, our method shows that the
action of the Heisenberg hamiltonian is the same as the laplacian
on the configuration-space graph of the hopping fermions. In this
way, as $N\rightarrow\infty$ we expect the statics and dynamics of
the Heisenberg and $XY$ models on families of regular graphs will
be very similar.

Finally, we point out that our method provides a very simple way
to qualitatively understand the quantum dynamics of $XY$ and
Heisenberg models on a graph $G$. The idea is that, for
intermediate time scales (i.e.\ time scales up to the order of
$N$), and reasonably well-separated spins, one can understand the
quantum dynamics as equivalent to the dynamics of a single free
particle hopping on the cartesian product graph $G\times G\times
\cdots \times G$.

Many future directions suggest themselves at this stage. The most
obvious direction is to study the spectral properties of the graph
wedge product in detail.

Another promising direction would be to investigate the
thermodynamic limit $N\rightarrow\infty$ of $H_{XY}$ and
$H_{\text{Heis}}$ on certain families of graphs $G_N$, such as the
path $P_N$, the cycle $C_N$, and cartesian products, which have a
well-defined continuum limit. In this limit the action of $H_{XY}$
and $H_{\text{Heis}}$ are simply related to the action of the
laplacian on the corresponding smooth manifold obeying certain
boundary conditions. Even in the mesoscopic limit of large but
finite $N$ we should be able to say something about how the
spectrum of $H_{XY}$ is related to $H_{\text{Heis}}$, potentially
providing a concrete analytical proof of the correctness of the
scaling hypotheses at criticality for these models.

A final future direction which presents itself is to investigate
the construction of models which have a gap in limit $N\rightarrow
\infty$. There are many examples of families of graphs which have
a spectral gap in the infinite limit.

\begin{acknowledgements}
I would like to thank Andreas Winter and Simone Severini for many
inspiring discussions. I am grateful to the EU for support for
this research under the IST project RESQ.
\end{acknowledgements}


\begin{thebibliography}{19}
\expandafter\ifx\csname
natexlab\endcsname\relax\def\natexlab#1{#1}\fi
\expandafter\ifx\csname bibnamefont\endcsname\relax
  \def\bibnamefont#1{#1}\fi
\expandafter\ifx\csname bibfnamefont\endcsname\relax
  \def\bibfnamefont#1{#1}\fi
\expandafter\ifx\csname citenamefont\endcsname\relax
  \def\citenamefont#1{#1}\fi
\expandafter\ifx\csname url\endcsname\relax
  \def\url#1{\texttt{#1}}\fi
\expandafter\ifx\csname urlprefix\endcsname\relax\def\urlprefix{URL
}\fi \providecommand{\bibinfo}[2]{#2}
\providecommand{\eprint}[2][]{\url{#2}}

\bibitem[{\citenamefont{Bethe}(1931)}]{bethe:1931a}
\bibinfo{author}{\bibfnamefont{H.~A.} \bibnamefont{Bethe}},
  \bibinfo{journal}{Z. Physik} \textbf{\bibinfo{volume}{71}},
  \bibinfo{pages}{205} (\bibinfo{year}{1931}).

\bibitem[{\citenamefont{Jordan and Wigner}(1928)}]{jordan:1928a}
\bibinfo{author}{\bibfnamefont{P.}~\bibnamefont{Jordan}} \bibnamefont{and}
  \bibinfo{author}{\bibfnamefont{E.}~\bibnamefont{Wigner}},
  \bibinfo{journal}{Z. Phys.} \textbf{\bibinfo{volume}{47}},
  \bibinfo{pages}{631} (\bibinfo{year}{1928}).

\bibitem[{\citenamefont{Fannes et~al.}(1994)\citenamefont{Fannes, Nachtergaele,
  and Werner}}]{fannes:1994a}
\bibinfo{author}{\bibfnamefont{M.}~\bibnamefont{Fannes}},
  \bibinfo{author}{\bibfnamefont{B.}~\bibnamefont{Nachtergaele}},
  \bibnamefont{and} \bibinfo{author}{\bibfnamefont{R.~F.}
  \bibnamefont{Werner}}, \bibinfo{journal}{J. Funct. Anal.}
  \textbf{\bibinfo{volume}{120}}, \bibinfo{pages}{511} (\bibinfo{year}{1994}),
  ISSN \bibinfo{issn}{0022-1236}.

\bibitem[{\citenamefont{Schollw{\"o}ck}(2005)}]{schollwoeck:2005a}
\bibinfo{author}{\bibfnamefont{U.}~\bibnamefont{Schollw{\"o}ck}},
  \bibinfo{journal}{Rev. Modern Phys.} \textbf{\bibinfo{volume}{77}},
  \bibinfo{pages}{259} (\bibinfo{year}{2005}), \eprint{cond-mat/0409292}.

\bibitem[{\citenamefont{Sachdev}(1999)}]{sachdev:1999a}
\bibinfo{author}{\bibfnamefont{S.}~\bibnamefont{Sachdev}},
  \emph{\bibinfo{title}{Quantum phase transitions}}
  (\bibinfo{publisher}{Cambridge University Press},
  \bibinfo{address}{Cambridge}, \bibinfo{year}{1999}).

\bibitem[{\citenamefont{Auerbach}(1994)}]{auerbach:1994a}
\bibinfo{author}{\bibfnamefont{A.}~\bibnamefont{Auerbach}},
  \emph{\bibinfo{title}{Interacting electrons and quantum magnetism}}
  (\bibinfo{publisher}{Springer-Verlag}, \bibinfo{address}{New York},
  \bibinfo{year}{1994}).

\bibitem[{\citenamefont{Biggs}(1993)}]{biggs:1993a}
\bibinfo{author}{\bibfnamefont{N.}~\bibnamefont{Biggs}},
  \emph{\bibinfo{title}{Algebraic graph theory}}, Cambridge Mathematical
  Library (\bibinfo{publisher}{Cambridge University Press},
  \bibinfo{address}{Cambridge}, \bibinfo{year}{1993}), \bibinfo{edition}{2nd}
  ed.

\bibitem[{\citenamefont{Cvetkovi{\'c} et~al.}(1995)\citenamefont{Cvetkovi{\'c},
  Doob, and Sachs}}]{cvetkovic:1995a}
\bibinfo{author}{\bibfnamefont{D.~M.} \bibnamefont{Cvetkovi{\'c}}},
  \bibinfo{author}{\bibfnamefont{M.}~\bibnamefont{Doob}}, \bibnamefont{and}
  \bibinfo{author}{\bibfnamefont{H.}~\bibnamefont{Sachs}},
  \emph{\bibinfo{title}{Spectra of graphs}} (\bibinfo{publisher}{Johann
  Ambrosius Barth}, \bibinfo{address}{Heidelberg}, \bibinfo{year}{1995}),
  \bibinfo{edition}{3rd} ed.

\bibitem[{\citenamefont{Chung}(1997)}]{chung:1997a}
\bibinfo{author}{\bibfnamefont{F.~R.~K.} \bibnamefont{Chung}},
  \emph{\bibinfo{title}{Spectral graph theory}}, vol.~\bibinfo{volume}{92} of
  \emph{\bibinfo{series}{CBMS Regional Conference Series in Mathematics}}
  (\bibinfo{publisher}{Published for the Conference Board of the Mathematical
  Sciences, Washington, DC}, \bibinfo{year}{1997}).

\bibitem[{\citenamefont{Lieb and Robinson}(1972)}]{lieb:1972a}
\bibinfo{author}{\bibfnamefont{E.~H.} \bibnamefont{Lieb}} \bibnamefont{and}
  \bibinfo{author}{\bibfnamefont{D.~W.} \bibnamefont{Robinson}},
  \bibinfo{journal}{Comm. Math. Phys.} \textbf{\bibinfo{volume}{28}},
  \bibinfo{pages}{251} (\bibinfo{year}{1972}).

\bibitem[{\citenamefont{Hastings}(2004)}]{hastings:2004a}
\bibinfo{author}{\bibfnamefont{M.~B.} \bibnamefont{Hastings}},
  \bibinfo{journal}{Phys. Rev. B} \textbf{\bibinfo{volume}{69}},
  \bibinfo{pages}{104431} (\bibinfo{year}{2004}), \eprint{cond-mat/0305505}.

\bibitem[{\citenamefont{Nachtergaele and Sims}(2005)}]{nachtergaele:2005a}
\bibinfo{author}{\bibfnamefont{B.}~\bibnamefont{Nachtergaele}}
  \bibnamefont{and} \bibinfo{author}{\bibfnamefont{R.}~\bibnamefont{Sims}}
  (\bibinfo{year}{2005}), \eprint{quant-ph/0506030}.

\bibitem[{\citenamefont{Hastings and Koma}(2005)}]{hastings:2005b}
\bibinfo{author}{\bibfnamefont{M.~B.} \bibnamefont{Hastings}} \bibnamefont{and}
  \bibinfo{author}{\bibfnamefont{T.}~\bibnamefont{Koma}}
  (\bibinfo{year}{2005}), \eprint{math-ph/0507008}.

\bibitem[{\citenamefont{Fulton and Harris}(1991)}]{fulton:1991a}
\bibinfo{author}{\bibfnamefont{W.}~\bibnamefont{Fulton}} \bibnamefont{and}
  \bibinfo{author}{\bibfnamefont{J.}~\bibnamefont{Harris}},
  \emph{\bibinfo{title}{Representation theory}}
  (\bibinfo{publisher}{Springer-Verlag}, \bibinfo{address}{New York},
  \bibinfo{year}{1991}).

\bibitem[{\citenamefont{Lang}(2002)}]{lang:2002a}
\bibinfo{author}{\bibfnamefont{S.}~\bibnamefont{Lang}},
  \emph{\bibinfo{title}{Algebra}}, vol. \bibinfo{volume}{211} of
  \emph{\bibinfo{series}{Graduate Texts in Mathematics}}
  (\bibinfo{publisher}{Springer-Verlag}, \bibinfo{address}{New York},
  \bibinfo{year}{2002}), \bibinfo{edition}{3rd} ed.

\bibitem[{\citenamefont{Alavi et~al.}(1991)\citenamefont{Alavi, Behzad,
  Erd{\H{o}}s, and Lick}}]{alavi:1991a}
\bibinfo{author}{\bibfnamefont{Y.}~\bibnamefont{Alavi}},
  \bibinfo{author}{\bibfnamefont{M.}~\bibnamefont{Behzad}},
  \bibinfo{author}{\bibfnamefont{P.}~\bibnamefont{Erd{\H{o}}s}},
  \bibnamefont{and} \bibinfo{author}{\bibfnamefont{D.~R.} \bibnamefont{Lick}},
  \bibinfo{journal}{J. Combin. Inform. System Sci.}
  \textbf{\bibinfo{volume}{16}}, \bibinfo{pages}{37} (\bibinfo{year}{1991}).

\bibitem[{\citenamefont{Alavi et~al.}(2002)\citenamefont{Alavi, Lick, and
  Liu}}]{alavi:2002a}
\bibinfo{author}{\bibfnamefont{Y.}~\bibnamefont{Alavi}},
  \bibinfo{author}{\bibfnamefont{D.~R.} \bibnamefont{Lick}}, \bibnamefont{and}
  \bibinfo{author}{\bibfnamefont{J.}~\bibnamefont{Liu}},
  \bibinfo{journal}{Graphs Combin.} \textbf{\bibinfo{volume}{18}},
  \bibinfo{pages}{709} (\bibinfo{year}{2002}), \bibinfo{note}{graph theory and
  discrete geometry (Manila, 2001)}.

\bibitem[{\citenamefont{Christandl et~al.}(2004)\citenamefont{Christandl,
  Datta, Ekert, and Landahl}}]{christandl:2003a}
\bibinfo{author}{\bibfnamefont{M.}~\bibnamefont{Christandl}},
  \bibinfo{author}{\bibfnamefont{N.}~\bibnamefont{Datta}},
  \bibinfo{author}{\bibfnamefont{A.}~\bibnamefont{Ekert}}, \bibnamefont{and}
  \bibinfo{author}{\bibfnamefont{A.~J.} \bibnamefont{Landahl}},
  \bibinfo{journal}{Phys. Rev. Lett.} \textbf{\bibinfo{volume}{92}},
  \bibinfo{pages}{187902} (\bibinfo{year}{2004}), \eprint{quant-ph/0309131}.

\bibitem[{\citenamefont{Godsil}(1995)}]{godsil:1995a}
\bibinfo{author}{\bibfnamefont{C.~D.} \bibnamefont{Godsil}}, in
  \emph{\bibinfo{booktitle}{Surveys in combinatorics, 1995 (Stirling)}}
  (\bibinfo{publisher}{Cambridge Univ. Press}, \bibinfo{address}{Cambridge},
  \bibinfo{year}{1995}), vol. \bibinfo{volume}{218} of
  \emph{\bibinfo{series}{London Math. Soc. Lecture Note Ser.}}, pp.
  \bibinfo{pages}{1--23}.

\bibitem{endnote15}{We write $|a_0,a_1,\protect \ldots
,a_{k-1}\delimiter "526930B $ to mean $|a_0\delimiter "526930B
\otimes |a_1\delimiter "526930B \otimes \protect \cdots  \otimes
|a_{k-1}\delimiter "526930B $.}

\end{thebibliography}
\end{document}